\documentclass[twocolumn,aps,prl,floatfix,amssymb,amsfonts,amsmath]{revtex4}

\usepackage{graphicx,graphics}
\usepackage{times}
\usepackage{psfrag}

\newcommand{\be}{\begin{equation}}
\newcommand{\ee}{\end{equation}}

\begin{document}

\preprint{Submitted to Phys. Rev. B}

%Title of paper
\title{
Coupling bosonic modes with a qubit: entanglement dynamics at zero
and finite-temperatures
 }

\author{Emanuele Ciancio}
\email[]{ciancio@isiosf.isi.it}
\author{Paolo Zanardi}
\email[]{zanardi@isiosf.isi.it}
\affiliation{
Institute for Scientific Interchange (ISI) \\
Viale Settimio Severo 65 \\
10133 Torino, Italy
}

\date{\today}

\begin{abstract}
We consider a system  of two iso-spectral bosonic modes coupled with
a single two-level systems i.e., a qubit. The dynamics is described
by a mode-symmetric  two-modes Jaynes-Cummings. The entanglement,
induced between the two bosonic modes, is analyzed and quantified by
negativity.
We computed the time evolution of
negativity  starting from an initial thermal state
of the bosonic sector for both zero and finite temperature.
We also studied  the entangling
power of the interaction as a function of mode-qubit detuning
and its resilience against temperature increase. Finally
a two-qubit gates based on bosonic virtual subsystem is discussed.
\end{abstract}

\maketitle

\section{Introduction}\label{s:intro}

In several of the existing implementation proposals for quantum
information processing \cite{qip} hybrid system  are involved. By
hybrid here we simply mean that both discrete i.e., atomic, and
continuous  e.g., vibrational modes in an ion-trap \cite{CZ},
degrees of freedom are present and interacting. Typically the former
are chosen to play the role of quantum  information carrier whereas
the latter are treated either  as information buses (to couple on demand
remote discrete systems e.g., cavity modes coupling atoms
\cite{QED1, QED2}) or are
hold responsible for decoherence effects e.g., phonons in quantum
dots. In either cases the attention is, in a sense,  mostly focused
on the  information-carrying discrete sector, and the bosonic modes
are eventually traced away.

In this paper we  would like, so to speak, to reverse this logic and to
address  the problem of
 bosonic modes interacting using  discrete systems  as interaction medium.
We will consider the  simplest instance of this situation: two
iso-spectral bosonic modes coupled through the interplay of one
qubit. We consider here the bosonic degrees of freedom as the
relevant ones for encoding information, while the qubit  degrees of
freedom  will be traced out. Information will be encoded in a
finite-dimensional subspace of the two-mode bosonic Fock space.

Of course the use of continuous variables i.e., light modes in the context
of quantum information processing has been already massively investigated
in the recent literature.
On one hand
there is the effort to handle with  the infinite dimensional Fock space of
bosonic degrees of freedom by mapping it onto finite dimensional ones by choosing ad
hoc states. On the other hand a
careful choice of a suitable subspace of the Fock space can also
provide  a  proper way for encoding  and processing qubits \cite{gpk}
The first approach is the one
contained e.g. in \cite{illu} in which the gaussian states are
involved in view of their key feature of being easily mappable
onto the
finite dimensional space covariance matrices. The second approach
instead aims at finding information-encoding states whose time evolutions remains inside
a finite dimensional subspace, at least for properly selected time instants
\cite{lars, WS, pater}.

This second approach is the one followed in
this paper. Given the interaction Hamiltonian, a proper encoding i.e., a two-dimensional Fock subspace,
is searched. The requirement that one can individuate  time
instants in which a non-trivial  unitary
transformation of the encoding subspace  is enacted.

More specifically, in this paper we shall analyze a symmetric
two-modes Jaynes-Cummings model. The entanglement properties of the
bosonic subsystem  are investigated by computing the dynamical evolution of
the negativity \cite{VW} of a generic state of the system.
Analytical and numerical results respectively are found for the
vacuum and the thermal initial state. The latter case is also
studied to prove the entanglement resilience against temperature and
qubit-modes detuning. The entanglement capabilities of our system
 are then
analyzed  by introducing a convenient notion  of entangling power
\cite{ep,giorda} and by observing its behaviour as a function of the
model parameters. Finally a unitary quantum gate for this kind of
interaction, is contrived. The realization of this gate however will
require the use of virtual subsystems \cite{virtual} and of the
related entanglement relativity concept \cite{cgz}.

\section{THE MODEL}

The physical system  studied in this paper is the one described by the two-mode
Jaynes-Cummings hamiltonian \cite{lars, WS}:

\begin{equation}
H = \sigma_z \epsilon + \sum_{i=1}^2 \hbar \omega_i b_i^{\dagger}b_i +
    \sum_{i=1}^2 \gamma(\sigma_- b_i^{\dagger} + \sigma_+ b_i)
\label{Ham}
\end{equation}

The symbols $b_i$ and $b_i^{\dagger}$ stand for annihilation and
creation of a particle (photon or phonon) in the mode $i$, while
$\gamma$ is the coupling constant and $\epsilon$ is the energy
difference between the two states. The hamiltonian simplifies
when i) the two modes have the same energy $\omega_1 = \omega_2 =
\omega$, and ii) the system achieves resonance $\epsilon =
\hbar\omega$. For the sake of simplicity we deal first with this
case. In order to diagonalize the hamiltonian we perform a unitary
transformation of the two mode operators, introducing two new modes:
$%\begin{eqnarray}
b_+ = \frac{1}{\sqrt{2}} (b_1 + b_2)
$ and $ %\nonumber\\
b_- = \frac{1}{\sqrt{2}} (b_1 - b_2)
$, %\end{eqnarray}
as well as
their conjugated $b_+^{\dagger}, b_-^{\dagger}$, and the inverse
equations:
$%\begin{eqnarray}
b_1 = \frac{1}{\sqrt{2}} (b_+ + b_-)
$ and $%\nonumber\\
b_2 = \frac{1}{\sqrt{2}} (b_+ - b_-)
$. %\end{eqnarray}
The hamiltonian then becomes:
\begin{equation}
H = \sigma_z \epsilon + \hbar \omega b_+^{\dagger}b_+ +
    \gamma(\sigma_- b_+^{\dagger} + \sigma_+ b_+) +
    \hbar \omega b_-^{\dagger}b_-
\end{equation}
As we can see, now the hamiltonian is the single-mode
Jaynes-Cummings hamiltonian plus a commuting term (the last one).
The interaction involves now only one mode while the other does not
evolve. The evolution of a generic (pure) state is now easy to
obtain following for example ref.\cite{JC, EP}; if the system is
prepared in the initial state with the qubit in its excited state
($e$) and the two bosonic modes occupied respectively with $n_1$ and
$n_2$ particles ($|\psi\rangle = |e,n_1,n_2\rangle = \sum_{n_+,n_-}
a_{n_+,n_-} |e,n_+,n_-\rangle$), the evolution will be:
%\begin{widetext}
\begin{eqnarray}\label{eqfond}
|\psi(t)\rangle &=& \sum_{n_1, n_2}
[%\left[
A_{n_1,n_2}(t)|e,n_1,n_2\rangle
\nonumber\\
&+& B_{n_1,n_2}(t)|g,n_1+1,n_2\rangle
+%\nonumber\\ &+&
C_{n_1,n_2}(t)|g,n_1,n_2 +1\rangle
]%\right] =
\nonumber\\
&=& \sum_{n_+,n_-} a_{n_+,n_-}
[%\left[
c_{n_+}(t)|e,n_+,n_-\rangle
\nonumber \\
&-& i s_{n_+}(t)|g,n_+ +1,n_-\rangle
]%\right]
\end{eqnarray}
%\end{widetext}
where $c_{n}(t) = \cos \Omega_{n} t, s_{n}(t) = \sin
\Omega_{n} t$ and $\Omega_n = \gamma \sqrt{n +1}$ is the (half) Rabi
frequency. The coefficients $a_{n_+,n_-}$ are obtained by writing
the initial state as a sum of states with fixed number of particles
in the new modes:
%\begin{widetext}
\begin{eqnarray}\label{psico1}
|\psi\rangle &=& |e,n_1,n_2\rangle = \frac{1}{\sqrt{n_1 ! n_2 !}}
         {b_1^{\dagger}}^{n_1} {b_2^{\dagger}}^{n_2} |e,0,0\rangle \nonumber\\
         &=&
         \left(\frac{1}{\sqrt{2}}\right)^{n_1 + n_2}
         \frac{1}{\sqrt{n_1 ! n_2 !}}
         (b_+^{\dagger} + b_-^{\dagger})^{n_1}
         (b_+^{\dagger} - b_-^{\dagger})^{n_2}
         |e,0,0\rangle \nonumber\\
         &=&
         \left(\frac{1}{\sqrt{2}}\right)^{n_1 + n_2}
         \frac{1}{\sqrt{n_1 ! n_2 !}}
         \times \nonumber\\ &\times&
         \sum_{i=0}^{n_1}
         \sum_{j=0}^{n_2} {n_1 \choose i} {n_2 \choose j} (-1)^j
         {b_+^{\dagger}}^{i+j} {b_-^{\dagger}}^{n_1 + n_2 -i -j} |e,0,0\rangle \nonumber\\
         &=&
         \sum_{n_+,n_-} a_{n_+,n_-} |e,n_+,n_-\rangle
\end{eqnarray}
%\end{widetext}
having identified, in the last passage, $i+j$ and $n_1 + n_2 -i -j$ with
$n_+$ and $n_-$ respectively.

With this formalism we can in principle describe the evolution of a
generic density operator:
%\begin{widetext}
\begin{eqnarray}\label{rotto1}
%\hskip -2mm
\rho &=& \sum_{n_1,n_2} p_{n_1,n_2} \rho_{n_1,n_2}
\nonumber\\&=&
\sum_{n_1,n_2} p_{n_1,n_2} |e,n_1,n_2\rangle\langle e,n_1,n_2|
=%\nonumber\\ &=&
\sum_{n_1,n_2} p_{n_1,n_2}
\times \nonumber\\
&\times&
\left[ \sum_{n_+,n_-} \sum_{n'_+,n'_-}
a_{n_+,n_-}a'_{n'_+,n'_-} |e,n_+,n_-\rangle\langle e,n'_+,n'_-|
\right]
\nonumber\\
\end{eqnarray}
%\end{widetext}
In general this straightforward procedure can be
quite difficult to handle because of the big amount of terms
involved in the sum. However in two cases the expression of the time
evolution of the initial state highly simplifies. They are the
vacuum and the thermal state cases.

\section{ENTANGLEMENT DYNAMICS}

In this section we will study the bosonic entanglement dynamics enacted by (\ref{Ham});
we will start from different initial preparations i.e., zero and finite temperature.
As entanglement measure we adopt  {\em negativity} introduced in Ref. \cite{VW}

\subsection{Vacuum State}

If the system is initially in the ground state with respect to the
two bosonic modes, it is straightforward to obtain: $\rho_{n_1,n_2}
= \rho_{n_+,n_-} = |e,0,0\rangle\langle e,0,0|$ being $n_i = n_{\pm} = 0$. The
time evolution of this state is given by:
\begin{eqnarray}
%\hskip -12mm
\rho(t) &=& c_0^2 (t) |e,0,0\rangle_{\pm}\langle e,0,0| + s_0^2 (t)
|g,1,0\rangle_{\pm}\langle g,1,0|\nonumber\\ %\nonumber\\
%\hskip -10mm
&+& i c_0 (t) s_0 (t) (|e,0,0\rangle_{\pm}\langle g,1,0| +
|g,1,0\rangle_{\pm}\langle e,0,0|) \nonumber\\ %\nonumber\\
\end{eqnarray}
Here and in the following we denote with $| \rangle_{12}$ and $|\rangle_{\pm}$
the elements of the two basis sets $ \{ |n_1\rangle \otimes |n_2\rangle \}$ and
$\{|n_+\rangle \otimes |n_-\rangle\}$ respectively.

We choose to use the negativity \cite{VW} as a  measure of
entanglement for bipartite states. The negativity is defined as the
sum of the negative eigenvalues of the partial transpose (transposed
with respect of one of the two subsystems) \cite{Peres, Hor} of the
density matrix of the state. In order to compute the negativity of
the state with respect to the bipartition defined by the two bosonic
modes we trace out the qubit degrees of freedom:
%\begin{widetext}
\begin{eqnarray}
%\nonumber\\
%& &
\rho(t) &=& c_0^2 (t) |0,0\rangle_{\pm}\langle0,0|
        + s_0^2 (t) b_+^{\dagger}|0,0\rangle_{\pm}\langle0,0|b_+ \nonumber\\&+&
        c_0^2 (t) |0,0\rangle_{12}\langle0,0| %\nonumber\\ &+&
        + \frac{s_0^2 (t)}{2} (b_1^{\dagger}+b_2^{\dagger})
                    |0,0\rangle_{12}\langle0,0|(b_1 + b_2) \nonumber\\
        &=& c_0^2 (t) |0,0\rangle_{12}\langle0,0| %\nonumber\\ &+&
        + \frac{s_0^2 (t)}{2} (|1,0\rangle_{12}\langle1,0| \nonumber\\
        &+& |1,0\rangle_{12}\langle0,1|
        +            |0,1\rangle_{12}\langle1,0| + |0,1\rangle_{12}\langle0,1|)
        %\nonumber\\
\end{eqnarray}
%\end{widetext}

The reduced density operator can be now represented as a matrix in
the $|n_1\rangle \otimes |n_2\rangle$ basis, and its partial transpose
$\rho^{T_1}$ with respect to (e.g.) the subsystem of mode $1$ reads:
\begin{eqnarray}
\rho^{T_1} = \left( \begin{array}{cccc}c_0^2 & 0 & 0 &
{\frac{1}{2}} s_0^2 \\
        0 & {\frac{1}{2}} s_0^2 & 0 & 0 \\
        0 & 0 & {\frac{1}{2}} s_0^2 & 0 \\
        {\frac{1}{2}} s_0^2 & 0 & 0 & 0\end{array}\right)
\end{eqnarray}

\begin{figure}[h]
\includegraphics[height=6cm, width=8cm, clip]{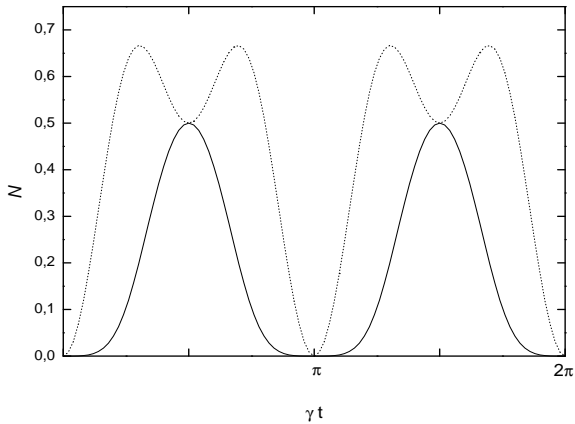}
\caption{Negativity as a function of time for the vacuum state
(solid line). The linearized Von Neumann entropy is plotted in dots
for comparison: the two lines touch each other when $\rho$ is a pure
state.} \label{tz}
\end{figure}

The negative part of the spectrum of this matrix gives the
analytical result for the negativity:
\begin{equation}
{\cal{N}} = | {1 \over 2} c_0^2 - {1 \over 2} \sqrt{c_0^4 + s_0^4} |
\end{equation}
whose graphical plot is displayed in fig.(\ref{tz}). It is
interesting to note the periodic behaviour of the negativity which
vanishes with period $\pi / \gamma$. On the other hand we can see
that this kind of interaction is able to entangle very much the
system, giving raise to maximally entangled states with the same
period $\pi / \gamma$. In fact as $t = \pi / 2\gamma +
K(\pi/\gamma)$ the state vector is (except for a phase factor):
\begin{equation}
|\psi\rangle = \frac{1}{\sqrt{2}} (|1,0\rangle_{12} + |0,1\rangle_{12})
\end{equation}
In fig.(\ref{tz}), besides the negativity, we plotted the linearized
Von Neumann entropy too: $S_L (\rho)= 1 - Tr_1 (Tr_2 \rho)^2$, with
$Tr_i$ meaning trace over the subsystem $i$. Its
analytical expressions for the present state is given by:
\begin{equation}
S_L = 1 - c_0^4 - {1 \over 2} s_0^4
\end{equation}
As long as $\rho(t)$ is a pure state, the Von Neumann entropy is a
 measure of entanglement and its maximum coincides with the
negativity of the maximally entangled state. The state $\rho(t)$ is
pure when $t = \pi / 2\gamma + K(\pi / 2\gamma)$, as it is easy to
recognize by computing its spectrum: $\sigma[\rho(t)] =
[0,0,c_0^2(t),s_0^2(t)]$. It is worth noting that when the state is
no more pure the Von Neumann entropy may exceed the maximum
entanglement value for pure states, then being a no more reliable
measure of entanglement, as fig.(\ref{tz}) shows.

%fig tzero

This result holds even for the more general case of a multi-mode
hamiltonian, and provides a quick way to generate the so-called $W$ n-qubit state
\begin{equation}
|\psi\rangle = \frac{1}{\sqrt{n}} (|1,...,0\rangle + ... + |0,...,1\rangle)
\end{equation}
This feature will be exploited in the design of a quantum gate in
the following.

\subsection{Thermal State}
Let us move now to the case in which both bosonic modes are in a thermal state.
The bosonic system will be described at the initial time as a tensor product
of the two thermal states:
\begin{equation}\label{biterm}
\rho = e^{-\beta \hbar\omega n_1} \otimes e^{-\beta \hbar\omega n_2}
     = e^{-\beta \hbar\omega (n_1 + n_2)}
     = e^{-\beta \hbar\omega (n_+ + n_-)}
\end{equation}
where, as usual, $\beta = \frac{1}{k_B T}$ and $n$ is used in the
operatorial meaning ($n = b^{\dagger}b$). The last identity of eq.
(\ref{biterm}) allows us to write the density operator directly in
the $|n_+\rangle \otimes |n_-\rangle$ basis:
$%\begin{eqnarray}
\rho
= %&=&
\sum_{n_1, n_2} p_{n_1,n_2} |n_1,n_2\rangle\langle n_1,n_2|
= %\nonumber\\ &=&
\sum_{n_+, n_-} p_{n_+,n_-} |n_+,n_-\rangle\langle n_+,n_-|
$ %\end{eqnarray}
where the sum goes as usual from zero to infinity. The coefficients
$p_n$ account for the Bose-Einstein probability distribution,
according to which:
\begin{equation}
p_n = \frac{\langle n\rangle^n}{(1+\langle n\rangle)^{n+1}}
\end{equation}
and the mean boson number is
\begin{equation}
\langle n\rangle = \frac{1}{e^{\beta \hbar \omega}-1}
\end{equation}
The time evolution of this thermal state, after tracing out the
qubit, will be:
\begin{eqnarray}
\rho(t) &=& \sum_{n_+,n_-} p_{n_+,n_-}
          %\left
          [c_{n_+}^2 (t) |n_+,n_-\rangle\langle n_+,n_-| + \nonumber\\
             &+& s_{n_+}^2 (t) |n_+ +1,n_-\rangle\langle n_+ +1,n_-|] % \right]
\end{eqnarray}

The matrix form of this operator, of course, is an infinite square
matrix whose spectrum can be computed numerically thanks to the
sparse structure of the matrix itself. By increasing the temperature
the higher energy levels begin to contribute to the density and the
infinite sum of terms has to be truncated carefully, taking into
account more and more levels. The matrix elements of
$\rho^{T_1}$ can be written as a sum of matrices with fixed number
of particles weighted with their thermal probability distribution.
By combining eq.(\ref{psico1}) and (\ref{rotto1}) as well as their
inverse, it is possible, after some tedious but straightforward
calculations, to write down the matrix elements of $\rho^{T_1}$ as:
\begin{widetext}
\begin{eqnarray}\label{megarho}
\hskip -2mm
\rho^{T_1}_{\mu\nu} &=& \lim_{N \rightarrow \infty}
\sum_{n =0}^N [\sum_{d_1 = 0}^{n} c_{d_1}^2(t) \frac{1}{2^n}
\sqrt{{d_1 \choose i} {d_1 \choose j}} \sqrt{{d_2 \choose k}{d_2
\choose l}} (-1)^{k+l} p_{d_1} p_{d_2} %\cdot \nonumber\\ \cdot
\delta_{\mu,n+1 +(N+2)(i+j)-(k+l)} \delta_{\nu,
n+1+(N+2)(k+l)-(i+j)} + \nonumber\\
%\hskip -8mm
&+& \sum_{d'_1 = 1}^{n+1} s_{d'_1 -1}^2(t) \frac{1}{2^n} \sqrt{{d'_1
\choose i} {d'_1 \choose j}} \sqrt{{d'_2 \choose k}{d'_2
\choose l}} (-1)^{k+l} p_{d'_1 -1} p_{d'_2} %\cdot \nonumber\\ \cdot
\delta_{\mu,n+2 +(N+2)(i+j)-(k+l)} \delta_{\nu,
n+2+(N+2)(k+l)-(i+j)} ]
\end{eqnarray}
\end{widetext}
where $d_2 = n - d_1$; $d'_2 = n+1- d'_1$ and $i,j = 0,...,d_1$;
$k,l = 0,...,d_2$.

\begin{figure}[h]
\includegraphics[height=7cm, width=8cm, clip]{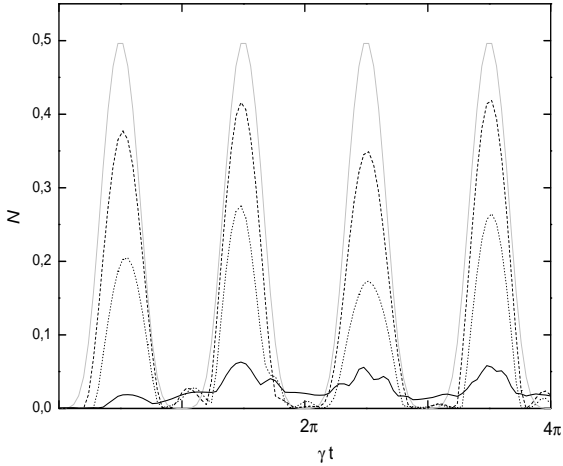}
\caption{The negativity behaviour of the thermal state as a function
of time is plotted for four different temperature. The latter are
labelled with different values of the parameter $1/\eta$: 0 for the
solid thin gray line, 0.5 for the dashed line, 1 for the dotted
line, 5 for the solid black line.} \label{negav}
\end{figure}

The matrix has a multi-diagonal form, meaning that the non-zero
elements are placed in diagonal lines parallel to principal diagonal
of the matrix. The computation of its spectrum and in particular of
its negative part (negativity) has been done in a numerical way. The
series appearing in eq.(\ref{megarho}) has been truncated using as a
convergence criterion: $\sum_{n=0}^{N} \sum_{d_1 =0}^n p_{d_1}
p_{d_2} \sim 1$. The results are shown in fig.(\ref{negav}) in which
the time evolution of the negativity is plotted for different values
of the energy/temperature ratio $\eta = \frac{\hbar\omega}{k_B T}$.
As one can expect, the negativity peaks get lower as temperature
raises. Nevertheless, surprisingly enough, the negativity remains
different from zero for long time even at high temperature. At $T=0$
we recover the same results of the previous subsection.

%fig neg

\begin{figure}[h]
\includegraphics[height=6cm, width=8cm, clip]{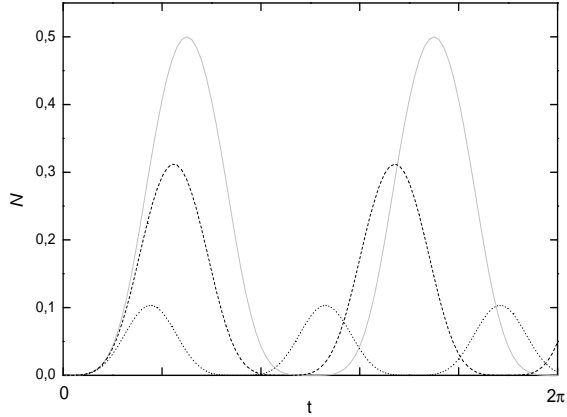}
\caption{Time variation of the negativity at $T=0$ for different
values of the detuning parameter: $\Delta = 0$ (thin gray line),
$\Delta=1$ (dashed line), $\Delta=2$ (dotted line).
%$\Delta=5$ (solid line).
Here the coupling constant is set as $\gamma =
1$.}\label{det0}
\end{figure}

\begin{figure}[h]
\includegraphics[height=7cm, width=9cm, clip]{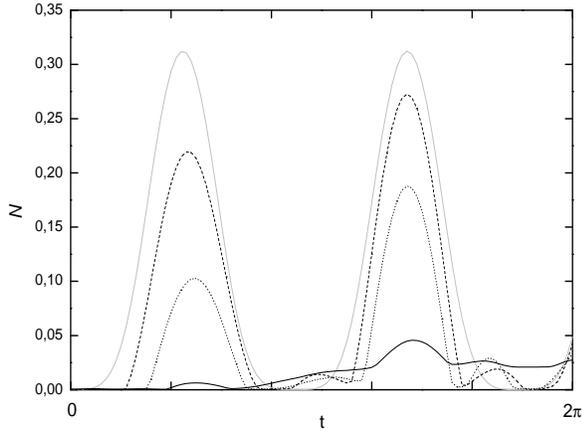}
\caption{Plot of the negativity time dependance in the case a fixed
detuning $\Delta=1$. Three different values of $1/\eta$ have been
chosen: 0 (thin gray line), 0.5 (dashed line), 1 (dotted line), 5
(solid line). As in the previous case: $\gamma=1$.}\label{detvar}
\end{figure}

In order to make our analysis slightly more general, we now extend
these results to the  case in which the qubit and the bosons are
off-resonance. This occurs when the qubit energy splitting and the
bosonic mode frequencies are different. In this situation the
coefficients $c_n$ and $s_n$ have a more complicated time dependence
\cite{JC}: $c_n = [\cos (\Omega_n t) - \frac{i\Delta}{2\Omega_n}
\sin (\Omega_n t)] \exp (i\Delta t/2)$ and $s_n =
(i\gamma\sqrt{n+1}/\Omega_n)\sin (\Omega_n t)\exp(i\Delta t/2)$,
while the (half) Rabi frequency becomes: $ \Omega_{n} = {1 \over
2}\sqrt{\Delta^2 + 4\gamma^2(n+1)}$, with $\Delta = (\epsilon -
\hbar\omega)/\hbar$. Fig.(\ref{det0}) shows the negativity behaviour
at $T=0$ for increasing values of $\Delta$, while fig.(\ref{detvar})
shows how the negativity varies in time for different temperatures
with a fixed non-vanishing value of the detuning ($\Delta=1$). As a
general remark, it is worth noting that an increasing detuning
causes the peaks of the negativity to lower at fixed temperature. In
a similar way the high temperatures lower the negativity as already
known from the resonance case (fig.(\ref{negav}) and (\ref{detvar})
are quite similar). Finally for high values of $\Delta$ the
negativity oscillates with frequency $\gamma^2/\Delta$ as a result
of a  second order effective interaction hamiltonian $H_{eff}$:
\begin{equation}\label{eff}
H_{eff} = \hbar\Omega_{eff} b_1^{\dagger} b_2 + h.c.
\end{equation}
where $\Omega_{eff} = \gamma^2/\Delta$. In fig.(\ref{det0}) the
dotted line plots the negativity for $\Delta = 2$ and $\gamma = 1$.
Since $\Delta >> \gamma$ we can consider $H_{eff}$ as a perturbation
and we can estimate the energies of the evolving states as the
unperturbed energy splitting $\epsilon \sim \Delta$ plus a second
order energy correction $\gamma^2/\Delta$. The same result is
obtained by considering the original hamiltonian with $ 2\Omega_{0}
= \sqrt{\Delta^2 + 4\gamma^2} = \Delta\sqrt{1 + 4\gamma^2/\Delta^2}
\sim \Delta [1+2(\gamma^2/\Delta^2)] = \Delta + 2\gamma^2/\Delta$.
The negativity oscillates with double frequency with respect to the
state vectors, since its time evolution is dictated by $s_0^2$ and
$c_0^2$. Therefore its approximated Rabi frequency will be
$2\Omega_0 =\Delta + 2\gamma^2/\Delta = 3$ as it is clearly shown in
fig.(\ref{det0}).

\section{ENTANGLING POWER}

\begin{figure}[h]
\includegraphics[height=6cm, width=9cm, clip]{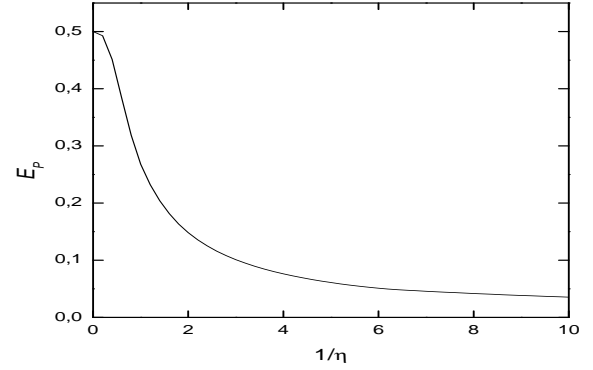}
\caption{The entangling power of the hamiltonian for the thermal
state as a function of the temperature. As long as $1/\eta$ is lower
than $1$, $E_P$ decreases very quickly, then it has a slow
variation, and for high values of the parameter it remains almost
constant.} \label{enpow}
\end{figure}

The above qualitative remarks have been quantitatively proven by
defining the entangling power \cite{giorda} of the hamiltonian as:
\begin{equation}
E_P (T) = \sup_{t\in \tau} {\cal{N}}\left[\rho_T(t)\right]
\end{equation}
where $\tau$ is the time variation period of $\rho$, and has been
chosen long enough to allow all the frequencies contributing the time
evolution to be taken into account. The reference state $\rho$
is a thermal state and the entangling power is computed as a
function of the temperature. Fig.(\ref{enpow}) shows its behaviour
for a range of the ratio $1/\eta$ going from $0$ to $10$. Although
the entangling power quickly decreases as temperature raises, it is
however different from zero for a wide range of temperatures.

%fig EP

\begin{figure}[h]
\includegraphics[height=5cm, width=8cm, clip]{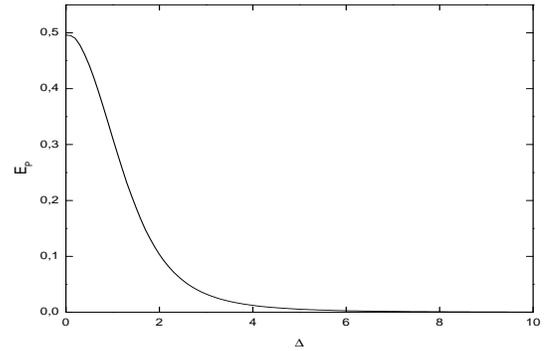}
\caption{Entangling power over all the off-resonant hamiltonian
labeled by the detuning parameter $\Delta$ going from 0 to 10.
}\label{epdet}
\end{figure}

As a second step we studied the entangling power of a wider set of
hamiltonian. As done in the previous section, we introduce the
detuning parameter $\Delta$ in the hamiltonian and redefine the
entangling power as:
\begin{equation}
E_P (\Delta) = \sup_{t\in \tau}
{\cal{N}}\left[\rho^{\Delta}(t)\right]
\end{equation}
where$\rho$ is now the vacuum state, chosen as reference state. The
result is displayed in fig.(\ref{epdet}) in which we see the
entangling power quickly decreasing down to zero, as $\Delta$
increases.
%Again this is easy to understand by looking at $H_{eff}$
%of eq.(\ref{eff}).

\begin{figure}[h]
\includegraphics[height=7cm, width=9cm, clip]{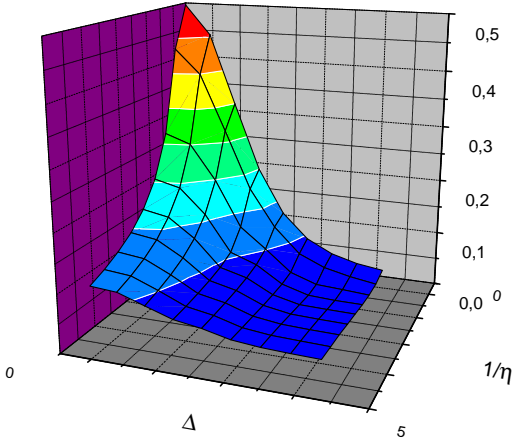}
\caption{Three-dimensional plot of the entangling power for both
$\Delta$ and $1/\eta$ going from 0 to 5. It is clear the monotone
character of $E_P$.}\label{ep3d}
\end{figure}

In order to combine the two previous results, we finally define the
entangling power as:
\begin{equation}
E_P (T,\Delta) = \sup_{t\in \tau}
{\cal{N}}\left[\rho^{\Delta}_T(t)\right]
\end{equation}
The reference state is again the thermal state, and the
3-dimensional plot of $E_P$ is represented in fig.(\ref{ep3d}). The
most interesting feature of $E_P$ is its monotonicity with respect
to both $T$ and $\Delta$.

\section{BUILDING A QUANTUM GATE}\label{qg}

%The above depicted analysis offers us the possibility to exploit the
%high entangling power of this two modes Jaynes-Cummings hamiltonian
%in order to design a quantum gate as a first step toward a realistic
%bosonic quantum computer.
In this section we will examine the possibility of using the above
analyzed entangling capabilities to build a quantum gate. The idea
is to encode  quantum information into the bosonic degrees of
freedom by selecting a suitable finite-dimensional subspace of the
Fock space. As mentioned in the introduction one would like to see
whether with the proper encoding one can find out specific operating
times such that the dynamics enacted by the Hamiltonian (\ref{Ham})
amounts to a non-trivial transformation of the encoding subspace
while acting  trivially in the qubit factor. This latter requirement
being of course due to the necessity of avoiding entanglement
between the bosons and the qubit that would in turn result into
decoherence.

 We consider as the encoded qubit the first two states
of harmonic oscillator of the modes of the system studied so far.
The logical computational basis is built by identifying: $|0\rangle_L =
|g,0\rangle_{12}$ and $|1\rangle_L =|g,1\rangle_{12}$. This is in a sense the most
natural basis choice. At variance with the previous calculations we
prepare the system with the qubit in its ground state \cite{EP}. The
two qubits evolution can be studied within the following scheme:
\begin{eqnarray}
|00\rangle_L &=& |g,0,0\rangle_{12} \nonumber\\
|10\rangle_L &=& |g,1,0\rangle_{12} \nonumber\\
|01\rangle_L &=& |g,0,1\rangle_{12} \nonumber\\
|11\rangle_L &=& |g,1,1\rangle_{12}
\end{eqnarray}

The time dependence of the four basis states is given by:
\begin{widetext}
\begin{eqnarray}\label{phsh}
|g,0,0\rangle_{12} &=& |g,0,0\rangle_{\pm} \rightarrow |g,0,0\rangle_{\pm} \nonumber\\
|g,1,0\rangle_{12} &=& \frac{1}{\sqrt{2}} (|g,1,0\rangle_{\pm} + |g,0,1\rangle_{\pm}) %\rightarrow \nonumber\\
\rightarrow \frac{1}{\sqrt{2}} (c_0 (t)|g,1,0\rangle_{\pm} - i s_0
(t)|e,0,0\rangle)_{\pm} +
               \frac{1}{\sqrt{2}} |g,0,1\rangle_{\pm} \nonumber\\
|g,0,1\rangle_{12} &=& \frac{1}{\sqrt{2}} (|g,1,0\rangle_{\pm} - |g,0,1\rangle_{\pm})
%\rightarrow \nonumber\\
\rightarrow \frac{1}{\sqrt{2}} (c_0 (t)|g,1,0\rangle_{\pm} - i s_0
(t)|e,0,0\rangle_{\pm}) -
               \frac{1}{\sqrt{2}} |g,0,1\rangle_{\pm} \nonumber\\
|g,1,1\rangle_{12} &=& \frac{1}{\sqrt{2}} (|g,2,0\rangle_{\pm} - |g,0,2\rangle_{\pm})
%\rightarrow \nonumber\\
\rightarrow
               \frac{1}{\sqrt{2}} (c_1 (t)|g,2,0\rangle_{\pm} - i s_1 (t)|e,1,0\rangle_{\pm}) -
               \frac{1}{\sqrt{2}} |g,0,2\rangle_{\pm} %\nonumber\\
\end{eqnarray}
\end{widetext}
With respect to the case in which the qubit was initially in its
excited state, in this case the global state of the system evolves
according to $c_{n_+ -1}$ and $s_{n_+ -1}$ instead of $c_{n_+}$ and
$s_{n_+}$.

%This feature indeed is the one that allows the last state in
%eq.(\ref{phsh}) to pick up a phase after $2\pi/\gamma$ units of
%time, while the remaining three states go back to the initial
%situation. In fact the first three states rotate with frequency
%$\gamma$ while the fourth one rotates with frequency
%$\sqrt{2}\gamma$, in such a way that after $2\pi/\gamma$ time units,
%its phase is shifted by $2\sqrt{2}\pi$.

Given this evolution table, it is possible to design a non-trivial
unitary transformation on the computational basis by properly tuning
the frequencies appearing int eq. (\ref{phsh}). By inspecting  at
the dynamical evolution of the basis it is clear that one can enact
the following gate:
\begin{eqnarray}
{U} = \left( \begin{array}{cccc} 1 & 0 & 0 & 0 \\
        0 & 0 & -1 & 0 \\
        0 & -1 & 0 & 0 \\
        0 & 0 & 0 & 1\end{array}\right)
\end{eqnarray}

Indeed, it is sufficient to find a time instant $t_g$ for which
$c_0(t_g)= -1, c_1(t_g) = 1$ and $s_0(t_g) = s_1(t_g) = 0$ to
realize the above unitary transformation. These conditions can be
achieved by choosing the variables $t_g,\gamma,\Delta$ such as the
following equation is fulfilled:
\begin{equation}
\frac{\Omega_0}{\Omega_1} = \frac{2N + 1}{2M}
\end{equation}
where$N,M$ are natural numbers, that means that for
$\Omega_0,\Omega_1$ we can choose $\forall q_0,q_1$ such that
$q_0/q_1 \in \mathbb{Q}$. Thus once we chose $q_0, q_1$, we get:
\begin{eqnarray}\label{param}
\gamma &=& \frac{1}{2} \sqrt{q_1^2 - q_0^2} \nonumber\\
\Delta &=& \sqrt {2 q_0^2 - q_1^2} \nonumber\\
t_g &=& \frac{\pi}{{1\over 2}q_0}(2N+1) = \frac{\pi}{{1\over
2}q_1}2M
\end{eqnarray}
whence we must add the conditions $q_1 > q_0$ and $2q_0^2 - q_1^2
\geq 0$ in order to have $\gamma, \Delta \in \mathbb{R}$. As an
example consider $q_0=\Omega_0=3$ and  $q_1=\Omega_1=4$; then
$4\gamma^2 = 7 , \Delta^2 = 2$ and $t_g = \pi + 2\pi K$.

The above sketched quantum gate however is the well known swap gate
together with a single qubit $\pi$-phase shift. Unfortunately it is
well known as well that this kind of gate, unless ancillary qubits
are introduced,
 is not an entangling one.
Anyway it is clear that no other unitary transformation is
realizable within the proposed scheme. In fact as soon as $t \neq 0$
the evolution brings all the basis states out of the computational
space, except for the first one. The fourth state comes back into
the computational space only when $c_1 = 1$ and $s_1 = 0$ giving
raise to the unchanged initial state. The same goes for the the
second and the third ($c_0 = 1$ and $s_1 = 0$), which have however
the extra possibility of getting (-)exchanged ($c_0 = -1$ and $s_0 =
0$). The situation would not change even if we wanted to start with
the qubit in its excited state. Again we would have to require that
all $c_n = e^{\pm i\phi}$ and $s_n = 0$, where the minus sign is
suitable only for the second and third state. Thus we could just
have the same phase shift for all the states. This is due to the
very fact that some of the coefficients (and so the frequencies)
appear in every basis state, preventing the latter to dephase one
from another. So one possibility to build a real quantum gate is to
choose a different computational basis which, although it may not be
scalable, allows however each state to evolve according to
frequencies different from the others.

We can in principle build a phase-shifter gate by mapping the
physical basis on the logical basis in a not straightforward way.
The previous exercise and eq. (\ref{eqfond}) revealed that in order
to have the basis last state's phase shifted alone, we need the
evolution frequencies of this state to be different from the other
states frequencies, at variance with what happens e.g. in eq.
(\ref{phsh}). A different choice of the basis would allows us to
overcome this obstacle. One should choose the computational basis
checking that:
%\begin{itemize}
%\item
i)
the states %of the $b_{\pm},b^\dagger_{\pm}$ modes
$|\psi\rangle_x = {1 \over N} (N_1|X,a_x,b_x\rangle_{\pm} + ... +
N_n|X,y_x,z_x\rangle_{\pm})$ have $a_x,b_x,...,y_x,z_x \neq
a_{x'},b_{x'},...,y_{x'},z_{x'}$, with $x,x' = 1,..,4$ and $X =
e,g$.
%\item
ii) one is able to find $t_g,\gamma,\Delta$ such that
$\Omega_{a_x},...,\Omega_{z_x}$ are commensurable frequencies.
%\end{itemize}
Unfortunately one has to deal with several commensurability conditions
that are not obvious to be satisfied.
%involving number theory problems are
%not easily realized, although in principle there is no theoretical
%obstacle in fulfilling them.
Therefore we will not further consider this procedure
any but we will focus on a second possibility, which is somehow
more elegant.

We can choose a virtual bipartition \cite{virtual} such that,
with respect to this new bipartition, the swap gate becomes an
entangling one. It is known indeed  that the
entanglement property of a state must be related to the choice of
the subsystems between which the quantum correlation is measured \cite{QTP,cgz}.
The swap operator itself can be an entangling one if we change the
subsystems (the qubits in our case), by redefining the modes and the
basis states. We know for example that the action of the swap
operator on the Bell basis states induces a sign change of the
antisymmetric state. Then we can change the bipartition in such a
way that the new subsystems (qubits) are no more the two oscillation
modes \cite{QTP}, but $|\chi\rangle$ and $|\lambda\rangle$ where $|\chi\rangle \otimes
|\lambda\rangle = |\chi^{\lambda}\rangle = |\Phi(\Psi)^{\pm}\rangle =
\frac{1}{\sqrt{2}}(|0,0(1)\rangle \pm |1,1(0)\rangle)$ with $\chi = \Phi,\Psi$
and $\lambda = +,-$. Thus by setting the computational basis as the
Bell one and identifying:
\begin{eqnarray}
|00\rangle_L &=& \frac{1}{\sqrt{2}}(|g,0,0\rangle_{12} + |g,1,1\rangle_{12}) \nonumber\\
|01\rangle_L &=& \frac{1}{\sqrt{2}}(|g,0,0\rangle_{12} - |g,1,1\rangle_{12}) \nonumber\\
|10\rangle_L &=& \frac{1}{\sqrt{2}}(|g,0,1\rangle_{12} + |g,1,0\rangle_{12}) \nonumber\\
|11\rangle_L &=& \frac{1}{\sqrt{2}}(|g,0,1\rangle_{12} - |g,1,0\rangle_{12})
\end{eqnarray}
we can let them evolve according to eqs. (\ref{phsh}) and
(\ref{param}), and get:
\begin{eqnarray}
{U'} = \left( \begin{array}{cccc} 1 & 0 & 0 & 0 \\
        0 & 1 & 0 & 0 \\
        0 & 0 & 1 & 0 \\
        0 & 0 & 0 & -1\end{array}\right)
\label{pi-shift}
\end{eqnarray}
%with the same parameters $t_g,\Delta,\gamma$ of the previous case .
Please note that, with respect to the previous case, now we are not
simply changing the basis, we are indeed changing the
computational subsystems. With this new choice the operator (\ref{pi-shift}) is a controlled $\pi$-phase
shift which can in principle be used for quantum computation
tasks.

It is important to stress that the use of virtual bi-partitions
makes now a non-trivial task the realization of single qubit gates
i.e., those with a non-trivial action on just one of the (virtual)
subsystems \cite{QTP}. On the other hand the aim of this section was
to discuss some issues related to gate building in the context of
the dynamics (\ref{Ham}) rather than proposing a universal set of
gates. For this latter more ambitious goal, it is well known that
some sort of non-linearity, possibly effective or even
measurement-induced, between the bosonic modes is necessary
\cite{KLM}.

\section{CONCLUSIONS}

In this paper we  have studied the behaviour of the negativity of
the quantum states of two iso-spectral bosonic modes symmetrically
coupled with a  common two-level system. By tracing over this latter
we showed that the interaction creates maximally entangled states
for $T=0$ in short time ($\pi / 2 \gamma $). We then turned to study
how entanglement production depends on the temperature by
considering as initial bosonic state the thermal one. Not
surprisingly the more the temperature raises the more entanglement
decreases. Nevertheless, we observed how entanglement resists even
up to relatively high $T$. A negativity peaks damping is also
observed as the result of a progressive increase of the detuning of
two-level system and the frequency of the bosonic modes: the more
the system departs  from the resonance condition i.e., detuning
zero, the lower maximally achievable negativity gets.

The analysis of the entangling power of the interaction suggested us
the way of designing a quantum gate adopting as computational basis
the degrees of freedom of the two bosonic subsystems (the modes).
The choice of the proper computational basis however turned up to be
not straightforward, involving the introduction of virtual
subsystems in order to be able to achieve a non-vanishing entangling
power. We would like now to conclude by briefly commenting upon
possible further investigations suggested by the study reported in
this paper.

It is clear that one  can consider the possibility of implementing
the theoretical framework discussed in the above with  systems
different from the light-matter interaction usually chosen as the
Jaynes-Cummings hamiltonian reference model. By properly adjusting
the parameters $\Delta, \gamma$ one might wonder whether, for
example, is possible to set up a controllable electron-phonon
interaction in a semiconductor heterostructure. Optical phonons can
interact via a Fr\"{o}lich coupling with electrons in the valence
band of a semiconductor superlattice designed to have only two
sublevels. With respect to the present model the latter would
involve more then one qubit i.e. atoms, of course. This would force
one to study a generalization of the simple model discussed above.
For instance one might have to consider some tailored periodic
potential profile instead of a single two-level step.

The study of electron-mediated quantum correlations between phononic
modes and  their potential usage of quantum-information carriers is
a particularly good illustration of the sort of logic which
motivated this paper (see introduction). Indeed phonons represent an
almost  prototypical example of an incoherent decoherence-inducing
system whereas electronic degrees of freedom are typically the ones
proposed for the encoding the quantum information \cite{LdV,biex}.
Phonons, typically in a thermal state, are traced out while ways to
enact coherent manipulations of electrons are contrived. In the
future we would like to investigate whether an exchange of these
roles of fermionic and bosonic degrees of freedom can be usefully
figured out.

%This kind of possible implementation can be a
%good motivation to study the quantum correlations of the phonons in
%the electronic bath, again reversing the common picture of the
%typical charge transport theory, where charge carriers (electrons)
%move in the phononic bath.

%\newpage

\end{document}